\documentclass{aastex}
\usepackage{spr-astr-addons}
\usepackage{url}

\RequirePackage{color}

\begin{document}

\title{Optical Photometry of GM Cep: Evidence for UXor Type of Variability}
\shorttitle{Optical Photometry of GM Cep}
\shortauthors{Semkov and Peneva}

\author{E.H. Semkov\altaffilmark{1}} \and \author{S.P. Peneva\altaffilmark{1}}

\email{esemkov@astro.bas.bg}

\altaffiltext{1}{Institute of Astronomy and National Astronomical Observatory, Bulgarian Academy of Sciences,
              72 Tsarigradsko Shose blvd., BG-1784 Sofia, Bulgaria}

\begin{abstract}

Results from optical photometric observations of the pre-main sequence star GM Cep are reported in the paper.
The star is located in the field of the young open cluster Trumpler 37 -- a region of active star formation.
GM Cep shows a large amplitude rapid variability interpreted as a possible outburst from EXor type in previous studies.
Our data from $BVRI$ CCD photometric observations of the star are collected from June 2008 to February 2011 in Rozhen observatory (Bulgaria) and Skinakas observatory (Crete, Greece).
A sequence of sixteen comparison stars in the field of GM Cep was calibrated in the $BVRI$ bands.
Our photometric data for a 2.5 years period show a high amplitude variations ($\Delta V$ $\sim$ $2\fm3$) and two deep minimums in brightness are observed.
The analysis of collected multicolor photometric data shows the typical of UX Ori variables a color reversal during the minimums in brightness.
On the other hand, high amplitude rapid variations in brightness typical for the Classical T Tauri stars also present on the light curve of GM Cep.
Comparing our results with results published in the literature, we conclude that changes in brightness are caused by superposition of both: (1) magnetically channeled accretion from the circumstellar disk, and (2) occultation from circumstellar clouds of dust or from features of a circumstellar disk.

\end{abstract}



\section{Introduction}

Photometric variability is a fundamental characteristic of the pre-main sequence (PMS) stars, which manifests as transient increases in brightness (outbursts), temporary drops in brightness (eclipses), irregular or regular variations for a short or long time scales.
The studies of photometric variability give us important information for the early stages of stellar evolution.
Both classes of PMS stars the widespread low-mass ($\it M$ $\leq$ $2M_\sun$) T Tauri Stars (TTSs) and the more massive Herbig Ae/Be Stars (HAEBESs) show various types of photometric variability (Herbst et al. 1994, 2007).
The TTSs can be separated into two subclasses: Classical T Tauri stars (CTT) surrounded by an extended circumstellar disk and Weak line T Tauri stars (WTT) without evidence of disk accretion (Bertout 1989).
Some authors (Herbst et al. 1994) consider appropriate to distinguish a third subclass composed by early type T Tauri stars (ETTS) with spectral class from K0 to A0.
Most of the stars in this subclass are actually HAEBES, and others are usually considered as CTTS (Herbst et al. 1994).
According to Herbst et al. (2007) the high amplitude variability of CTT stars is caused by magnetically channeled accretion from the circumstellar disk onto the star.
In this case the accretion is highly variable in time and the accretion zones are not uniformly distributed on the stellar surface.
The variations are most often irregular with amplitudes reaching up to $1\fm5$ ($V$) within a few days.

The observed large amplitude variations in brightness bring valuable information for the circumstellar environment and the circumstellar disks.
The large amplitude outbursts of PMS stars can be grouped into two main types, named after their respective prototypes: FU Orionis (FUor; Ambartsumian 1971) and EX Lupi (EXor; Herbig 1989). 
Both types of stars seem related to low-mass TTS with massive circumstellar disks, and their outbursts are generally attributed to a sizable increase in the disc accretion rate onto the stellar surface (Hartmann \& Kenyon 1996).
During the quiescence state FUors and EXors are normally accreting TTSs, but because of thermal instability in the circumstellar disk accretion rate enhanced by a few orders of magnitude from $\sim$10$^{-7}$$M_{\sun}$$/$yr up to $\sim$10$^{-4}$$M_{\sun}$$/$yr.
FUor objects are characterized by $\Delta$$V$$\approx$4-5 mag outburst amplitude, an F-G super-giant spectrum during outbursts, association with reflection nebulae, location in star-forming regions, a strong LiI~6707~\AA\ in absorption, H$\alpha$ and Na I 5890, 5896 \AA\ displaying P-Cyg profiles (Herbig 1977; Reipurth \& Aspin 2010). 
The outburst of FUor objects last for several decades, and the rise time is faster than the decline.
EXor objects shows frequent (every few years or decade), irregular and relatively brief (a few months to a few years) outburst of several magnitudes amplitude {\bf ($\Delta$$V$$\approx$3-5)}. 
During such events, the cool spectrum of the quiescence is veiled, and strong emission lines from single ionized metals are observed together with appearance of reversed P-Cyg absorption components (Herbig 2007).

A significant part of HAEBE stars with a spectral type later than A0 show strong photometric variability with sudden quasi-Algol drops in brightness and amplitudes up to $2\fm5$ ($V$) (Natta et al. 1997, van den Ancker et al. 1998). 
During the deep minimums of brightness, an increase in polarization and specific color variability are observed. 
The prototype of this group of PMS objects with intermediate mass named UXors is UX Orionis. 
The general explanation of its variability is a variable extinction from dust clumps of filaments passing through the line of sight (Grinin et al. 1991, Dullemond et al. 2003).

The variable star GM Cep was discovered seven decades ago on the photographic plates from Sonneberg observatory (Morgenroth 1939), but the exact mechanism of variability remains still under discussion (Xiao et al. 2010).  
GM Cep lie in the field of the young open cluster Trumpler 37 ($\sim$4 Myr old) and most likely is a member of the cluster (Marschall \& van Altena 1987, Sicilia-Aguilar et al. 2005).
According to Sicilia-Aguilar et al. (2008) GM Cep is a solar mass-star ($\it M$ $\sim$ $2.1 M_\sun$) from G7V-K0V spectral type, with radius between 3 and 6 $R_\sun$ and with a strong IR excesses from a very luminous circumstellar disk.  
The physical parameters (mass, radius, spectral type) of GM Cep defined it as member of the subclass of ETTS.
The spectrum of GM Cep is dominated by a very strong and broad H$\alpha$ emission line with a strong P Cyg profile (Sicilia-Aguilar et al. 2008). 
The equivalent width of H$\alpha$ line vary significantly from 6$\AA$ (2001 Jun.) to 19$\AA$ (2007 Apr.). 
But the lack of photometric observations around these dates do impossible to determine a correlation between the brightness of the star and the equivalent width of H$\alpha$ line. 
The presence of a massive circumstellar disk and variable accretion rates (up to $\sim$ 10$^{-6}$ $M_\sun$/year) are also detected in the study of Sicilia-Aguilar et al. (2008).

On the basis of photographic monitoring Suyarkova (1975) classified GM Cep as a RW Aur variable (also known as extreme CTTSs). 
The registered photographic amplitude for a nine years period is 2.2 mag.
Similar brightness variations are reported by Kun (1986) (the observed amplitude is $\Delta V$=$2\fm15$).
The first multicolor photometric study of GM Cep based on optical and infrared observations was done by Sicilia-Aguilar et al. (2008).
The authors found the star much brighter in 2006 than in 2000 and conclude that the most probable reason for brightness increase is an outburst from EXor type.
Sicilia-Aguilar et al. (2008) argue their hypothesis with the presence of luminous mid-IR disk and with the high and variable accretion rate.
A long-term photometric study of GM Cep for several decades period was performed by Xiao et al. (2010).
The photographic plate archives from Harvard College Observatory and from Sonneberg Observatory are used to construct the long-term $B$ and $V$ light  curves of the star.
The results suggest that GM Cep do not show fast rises in brightness typical of EXor variables and the light curves seem to be dominated by dips superposed on the quiescence state.
Evidences for periodicity of observed dips in brightness were not found.

Recent $BVRI$ CCD photometric observations of GM Cep are reported in the present paper.
We try to collect regular observations of the star in order to clarify the nature of this object.
The multicolor observations give us the opportunity to determine the mechanism of the brightness variations.

\section{Observations}

Our photometric CCD data were obtained in two observatories with three telescopes: the 2-m Ritchey-Chr\'{e}tien-Coud\'{e} and the 50/70-cm Schmidt telescopes of the National Astronomical Observatory Rozhen (Bulgaria) and the 1.3-m Ritchey-Cr\'{e}tien telescope of the Skinakas Observatory\footnote{Skinakas Observatory is a collaborative project of the University of Crete, the Foundation for Research and Technology - Hellas, and the Max-Planck-Institut f\"{u}r Extraterrestrische Physik.} of the Institute of Astronomy, University of Crete (Greece).

Observations were made with tree types of CCD camera - Vers Array 1300B at the 2-m RCC telescope, ANDOR DZ436-BV at the 1.3-m RC telescope, and FLI PL16803 CCD camera at the 50/70-cm Schmidt telescope.
The technical parameters and chip specifications for the cameras used are summarized in Table 1.
All frames were taken through a standard Johnson-Cousins set of filters.
Twilight flat fields in each filter were obtained each clear evening.
All frames obtained with the ANDOR and Vers Array cameras are bias subtracted and flat fielded.
CCD frames obtained with the FLI PL16803 camera are dark subtracted and flat fielded.
Aperture photometry was performed using DAOPHOT routines.

\begin{table*}
\small
\caption{CCD cameras and chip specifications}
\label{tablabel}
 \begin{tabular}{llllll}
            \hline
            \noalign{\smallskip}  Telescope & CCD type & Size & Pixel size & Field & RON \\
\noalign{\smallskip}
\hline
\noalign{\smallskip}
2-m RCC & VersArrey 1300B & $1340\times1300$ & 20 $\mu$m & $5.\mkern-4mu^\prime6\times5.\mkern-4mu^\prime6$  & 2.8 ADU/rms \\
1.3-m RC & ANDOR DZ436-BV& $2048\times2048$ & 13.5 $\mu$m & $9.\mkern-4mu^\prime6\times9.\mkern-4mu^\prime6$ & 5.3 ADU/rms\\
Schmidt & FLI PL16803    & $4096\times4096$ & 9 $\mu$m & $74^\prime\times74^\prime$ & 9 ADU/rms \\
\noalign{\smallskip}
            \hline
         \end{tabular}
   \end{table*}

\begin{figure}[tb]
\includegraphics[width=\columnwidth]{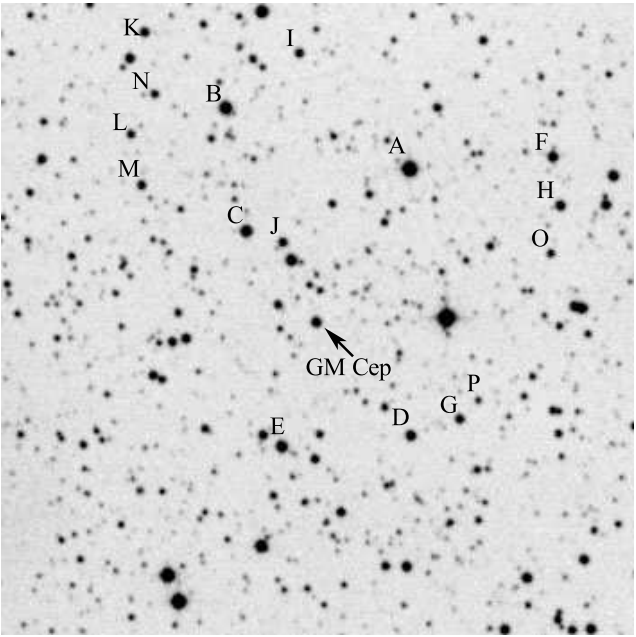}
\caption{A finding chart for the $\it BVRI$ comparison sequence around GM Cep} 
\end{figure}

In order to facilitate transformation from instrumental measurement to the standard Johnson-Cousins system sixteen stars in the field of GM Cep were calibrated in $BVRI$ bands.
Calibration was made during eight clear nights in 2008 and 2009 with the 1.3-m RC telescope of Skinakas Observatory. 
Standard stars from Landolt (1992) were used as reference.
The finding chart of the comparison sequence is presented in Fig. 1.
The comparison stars are labeled from A to P in order of their $V$ magnitudes.
The field is $6\arcmin\times6\arcmin$, centered on GM Cep.
North is at the top and east to the left.
The chart is retrieved from the STScI Digitized Sky Survey Second Generation Red.
Table 2 contains the co-ordinates and the photometric data for the $BVRI$ comparison sequence. 
The corresponding mean errors of the mean are listed, too.

\begin{table*}
\small
\caption{Photometric data for $\it BVRI $ comparison sequence}
\label{tablabel}
 \begin{tabular}{lllllllllll}

            \hline
            \noalign{\smallskip}          
Star & R.A. (2000) & DEC. (2000) & $\it B$ & $\sigma_B$& $\it V$ & $\sigma_V$ & $\it R$ & $\sigma_R$ &
$\it I$ & $\sigma_I $ \\
            \noalign{\smallskip}
            \hline
            \noalign{\smallskip}
A	&	21:38:11.10	&	57:32:50.9	&	12.39	&	0.02	&	11.77	&	0.02	&	11.40	&	0.01	&	11.06	&	0.01	\\
B	&	21:38:24.22	&	57:33:23.6	&	13.40	&	0.03	&	12.84	&	0.02	&	12.53	&	0.02	&	12.15	&	0.02	\\
C	&	21:38:22.41	&	57:32:13.6	&	16.32	&	0.02	&	14.25	&	0.03	&	13.01	&	0.04	&	11.73	&	0.08	\\
D	&	21:38:10.34	&	57:30:19.5	&	15.35	&	0.02	&	14.37	&	0.01	&	13.78	&	0.02	&	13.19	&	0.02	\\
E	&	21:38:19.48	&	57:30:10.6	&	16.67	&	0.03	&	14.39	&	0.04	&	13.00	&	0.04	&	11.62	&	0.09	\\
F	&	21:38:00.92	&	57:33:00.8	&	17.04	&	0.04	&	14.97	&	0.03	&	13.71	&	0.03	&	12.50	&	0.08	\\
G	&	21:38:06.94	&	57:30:30.0	&	16.03	&	0.03	&	14.98	&	0.02	&	14.37	&	0.02	&	13.77	&	0.03	\\
H	&	21:38:00.30	&	57:32:33.2	&	16.86	&	0.03	&	15.07	&	0.03	&	13.99	&	0.03	&	12.94	&	0.06	\\
I	&	21:38:19.05	&	57:33:55.5	&	15.82	&	0.03	&	15.08	&	0.02	&	14.63	&	0.02	&	14.12	&	0.03	\\
J	&	21:38:19.76	&	57:32:07.5	&	16.20	&	0.03	&	15.23	&	0.02	&	14.64	&	0.02	&	14.07	&	0.02	\\
K	&	21:38:30.06	&	57:34:04.8	&	17.03	&	0.03	&	15.37	&	0.03	&	14.40	&	0.02	&	13.44	&	0.06	\\
L	&	21:38:30.69	&	57:33:05.9	&	16.38	&	0.02	&	15.53	&	0.02	&	15.03	&	0.02	&	14.50	&	0.04	\\
M	&	21:38:29.88	&	57:32:37.7	&	16.87	&	0.04	&	15.57	&	0.02	&	14.69	&	0.02	&	13.99	&	0.04	\\
N	&	21:38:29.17	&	57:33:29.5	&	16.69	&	0.03	&	15.70	&	0.02	&	15.12	&	0.03	&	14.50	&	0.04	\\
O	&	21:38:00.88	&	57:32:05.7	&	17.03	&	0.03	&	15.93	&	0.02	&	15.28	&	0.02	&	14.60	&	0.04	\\
P	&	21:38:05.60	&	57:30:40.7	&	18.93	&	0.12	&	17.16	&	0.05	&	16.10	&	0.02	&	15.10	&	0.09	\\
            \noalign{\smallskip}
            \hline
         \end{tabular}
   \end{table*}

A sequence of comparison stars in the field around GM Cep was reported in Xiao et al. (2010) also.
The authors measured seven stars in $BVR$ Johnson-Cousins system using the 35-cm Schmidt-Cassegrain telescope of the Sonoita Research Observatory (USA).
Four of these stars were selected for comparisons and included in our list. 
The stars B, C, D and G from Table 2 correspond to stars G, A, D and B from the list of Xiao et al. (2010) respectively.
The comparison between our magnitudes and the measured by Xiao et al. (2010) shows good agreement of both sequences.
Differences in most cases are smaller than the errors of measurement and shall not exceed $0\fm04$.

The CCD photometric $BVRI$ data presented in this paper were collected from June 2008 to February 2011.
The results of our photometric observations of GM Cep are summarized in Table 3.
The columns give: Julian date, $BVRI$ magnitudes and telescope used.
At least two frames were taken in each filter every night, and then the average values are printed in the table.
Because the relatively brightness of GM Cep and the large number of comparison stars our photometric data were obtained with very good accuracy.
The values of instrumental errors are in the range $0\fm01$-$0\fm02$ (for $I$ and $R$), $0\fm01$-$0\fm03$ (for $V$) and $0\fm01$-$0\fm05$ (for $B$).
The $BVRI$-light curves of GM Cep during the period of our observations are plotted in Fig. 2.
In most cases (except a few $B$ points) the size of error bars are smaller than the size of the symbols used.

\begin{figure*} 
\includegraphics[width=17cm]{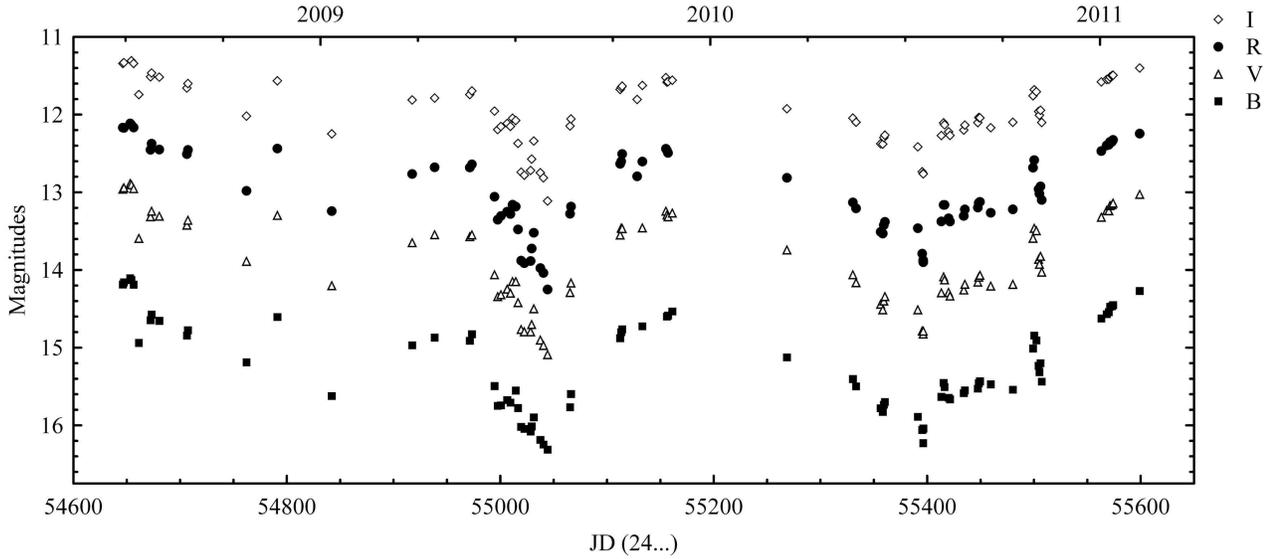}
\caption{$BVRI$ light curves of GM Cep in the period June 2008 -- February 2011}
\label{figlabel}
\end{figure*}

   \begin{table*}
   \small
   \caption[]{$BVRI$ photometric observations of GM Cep}
   \begin{tabular}{llllllllllll}
            \hline
            \noalign{\smallskip}
      JD (24...) & $\it B$  & $\it V$ & $\it R$ & $\it I$  &  Telescope & JD (24...) & $\it B$  & $\it V$ & $\it R$ & $\it I$  &  Telescope\\
            \noalign{\smallskip}
            \hline
            \noalign{\smallskip}
54646.43	&	14.19	&	12.96	&	12.17	&	11.34	&	1.3-m  	&	55155.21	&	   -	&	13.24	&	12.44	&	11.52	&	Schmidt	\\
54647.43	&	14.16	&	12.94	&	12.17	&	11.33	&	1.3-m	  &	55156.22	&	14.60	&	13.29	&	12.46	&	11.58	&	Schmidt	\\
54653.37	&	14.11	&	12.88	&	12.11	&	   -	&	1.3-m  	&	55157.25	&	14.59	&	13.31	&	12.49	&	11.58	&	Schmidt	\\
54654.43	&	14.13	&	12.90	&	12.13	&	11.31	&	1.3-m  	&	55161.22	&	14.54	&	13.27	&	   -	&	11.56	&	2-m	\\
54656.46	&	14.19	&	12.95	&	12.17	&	11.34	&	1.3-m  	&	55268.60	&	15.13	&	13.74	&	12.82	&	11.93	&	2-m	\\
54661.43	&	14.94	&	13.59	&	   -	&	11.74	&	1.3-m  	&	55330.41	&	15.41	&	14.06	&	13.13	&	12.05	&	Schmidt	\\
54672.34	&	14.65	&	13.31	&	12.45	&	11.51	&	1.3-m  	&	55333.39	&	15.50	&	14.16	&	13.21	&	12.10	&	Schmidt	\\
54673.36	&	14.58	&	13.24	&	12.37	&	11.46	&	1.3-m  	&	55356.39	&	15.78	&	14.44	&	13.51	&	12.38	&	Schmidt	\\
54680.53	&	14.66	&	13.31	&	12.45	&	11.52	&	1.3-m 	&	55358.45	&	15.83	&	14.51	&	13.53	&	12.38	&	Schmidt	\\
54706.29	&	14.85	&	13.42	&	12.51	&	11.66	&	Schmidt	&	55359.49	&	15.74	&	14.40	&	13.42	&	12.30	&	Schmidt	\\
54707.35	&	14.78	&	13.36	&	12.45	&	11.60	&	Schmidt	&	55360.43	&	15.70	&	14.34	&	13.38	&	12.27	&	Schmidt	\\
54762.23	&	15.19	&	13.89	&	12.98	&	12.02	&	Schmidt	&	55391.34	&	15.89	&	14.51	&	13.46	&	12.42	&	2-m	\\
54791.24	&	14.61	&	13.30	&	12.44	&	11.57	&	Schmidt	&	55395.34	&	16.06	&	14.78	&	13.79	&	12.74	&	2-m	\\
54842.20	&	15.62	&	14.20	&	13.24	&	12.25	&	Schmidt	&	55396.32	&	16.23	&	14.82	&	13.88	&	12.76	&	2-m	\\
54917.54	&	14.97	&	13.65	&	12.76	&	11.81	&	Schmidt	&	55396.36	&	16.04	&	14.78	&	13.90	&	12.76	&	Schmidt	\\
54938.47	&	14.87	&	13.54	&	12.68	&	11.79	&	Schmidt	&	55413.29	&	15.63	&	14.29	&	13.38	&	12.27	&	Schmidt	\\
54971.40	&	14.91	&	13.57	&	12.68	&	11.74	&	Schmidt	&	55415.38	&	15.45	&	14.09	&	13.16	&	12.11	&	Schmidt	\\
54973.43	&	14.83	&	13.55	&	12.64	&	11.70	&	Schmidt	&	55416.33	&	15.51	&	14.13	&	13.16	&	12.13	&	Schmidt	\\
54994.58	&	15.49	&	14.06	&	13.06	&	11.96	&	1.3-m  	&	55420.05	&	15.65	&	14.29	&	13.33	&	12.22	&	1.3-m	\\
54997.53	&	15.75	&	14.34	&	13.35	&	12.19	&	1.3-m  	&	55421.37	&	15.67	&	14.33	&	13.38	&	12.27	&	1.3-m	\\
55000.58	&	15.74	&	14.32	&	13.31	&	12.15	&	1.3-m  	&	55434.32	&	15.59	&	14.26	&	13.31	&	12.20	&	1.3-m	\\
55006.52	&	15.68	&	14.25	&	13.25	&	12.12	&	1.3-m	  &	55435.35	&	15.55	&	14.18	&	13.22	&	12.14	&	1.3-m	\\
55009.59	&	15.71	&	14.30	&	13.28	&	12.15	&	1.3-m  	&	55447.49	&	15.53	&	14.15	&	13.20	&	12.10	&	Schmidt	\\
55011.50	&	   -	&	14.15	&	13.16	&	12.05	&	Schmidt	&	55448.37	&	15.46	&	14.09	&	13.14	&	12.04	&	Schmidt	\\
55014.52	&	15.55	&	14.15	&	13.19	&	12.08	&	1.3-m	  &	55449.44	&	15.43	&	14.06	&	13.12	&	12.04	&	Schmidt	\\
55016.58	&	15.78	&	14.42	&	13.48	&	12.37	&	1.3-m	  &	55459.53	&	15.47	&	14.20	&	13.27	&	12.17	&	1.3-m	\\
55019.51	&	16.02	&	14.76	&	13.88	&	12.74	&	1.3-m	  &	55480.34	&	15.54	&	14.18	&	13.22	&	12.10	&	1.3-m	\\
55022.52	&	16.05	&	14.79	&	13.91	&	12.78	&	1.3-m	  &	55499.30	&	15.01	&	13.59	&	12.68	&	11.76	&	2-m	\\
55028.41	&	16.08	&	14.79	&	13.89	&	12.72	&	Schmidt	&	55500.24	&	14.85	&	13.46	&	12.59	&	11.68	&	2-m	\\
55029.42	&	16.01	&	14.70	&	13.73	&	12.57	&	Schmidt	&	55502.26	&	14.91	&	13.49	&	12.60	&	11.71	&	2-m	\\
55031.50	&	15.90	&	14.50	&	13.52	&	12.34	&	1.3-m	  &	55504.40	&	15.24	&	13.86	&	12.96	&	11.96	&	Schmidt	\\
55037.53	&	16.19	&	14.90	&	13.98	&	12.75	&	1.3-m	  &	55505.26	&	15.32	&	13.93	&	13.02	&	12.01	&	Schmidt	\\
55040.43	&	16.25	&	14.97	&	14.04	&	12.82	&	1.3-m	  &	55506.27	&	15.20	&	13.82	&	12.93	&	11.95	&	Schmidt	\\
55044.37	&	16.31	&	15.09	&	14.25	&	13.11	&	1.3-m	  &	55507.42	&	15.44	&	14.03	&	13.10	&	12.10	&	Schmidt	\\
55065.33	&	15.77	&	14.29	&	13.28	&	12.15	&	Schmidt	&	55563.20	&	14.63	&	13.32	&	12.47	&	11.58	&	Schmidt	\\
55066.26	&	15.60	&	14.17	&	13.19	&	12.06	&	Schmidt	&	55568.22	&	14.57	&	13.23	&	12.40	&	11.55	&	2-m	\\
55112.37	&	14.88	&	13.55	&	12.64	&	11.68	&	Schmidt	&	55570.22	&	14.54	&	13.24	&	12.39	&	11.55	&	2-m	\\
55113.33	&	14.80	&	13.46	&	12.60	&	11.65	&	Schmidt	&	55571.25	&	14.47	&	13.17	&	12.35	&	11.52	&	2-m	\\
55114.27	&	14.77	&	13.47	&	12.51	&	11.63	&	Schmidt	&	55573.25	&	14.47	&	13.16	&	12.35	&	11.50	&	2-m	\\
55128.28	&	   -	&	-	    &	12.79	&	11.81	&	2-m	    &	55574.26	&	14.45	&	13.14	&	12.33	&	11.49	&	2-m	\\
55133.26	&	14.73	&	13.46	&	12.61	&	11.63	&	Schmidt	&	55599.21	&	14.27	&	13.02	&	12.24	&	11.40	&	Schmidt	\\
\noalign{\smallskip}
            \hline
         \end{tabular}
   \end{table*}

\section{Results and Discussion}

Our observations confirm the strong brightness variability of GM Cep reported in the previous papers (Sicilia-Aguilar et al. 2008; Xiao et al. 2010).
The present photometric data show a large amplitude variations ($\Delta V\sim2\fm3$) during the 2.5 years period of observations. 
Most of the time the brightness of the star remains in position around the maximum, but two drops in brightness for short time intervals are observed.
The first deep minimum is registered at July-August 2009 and the second at July 2010 (Fig. 2).
The time interval between the two minimums is 350 days approximately.
Out of deep minimums GM Cep shows significant brightness variations in the time scale of days and months.
The fastest change in brightness was registered in November 2010, when for a seven days period the stellar magnitude decrease by $0\fm57 (V)$. 
A similar result can be found in the photometric data published by Sicilia-Aguilar et al. (2008). 
The fastest decrease in brightness with $1\fm019 (V)$ amplitude was registered during a fourteen days period (15-29 December 2006) followed by $0\fm816 (V)$ increase in brightness over a period of sixteen days (2-18 January 2007).

Such type of photometric variability is seen on the long-term light curves of GM Cep constructed by Xiao et al. (2010).
The authors indicate that the variability of GM Cep is more likely due to dips caused by extinction instead of outbursts caused by accretion.
The search for periodicity performed by Xiao et al. (2010) show no significant periods in the range between 0.5 and 100 days.
But periods of longer duration are not excluded and the collection of a long set of homogeneous multicolor observations can be used in a new search for periodicity.

Using our photometric data the color-magnitude diagrams of GM Cep are constructed and presented on Fig. 3.
The values of the three color indices $B-V$, $V-R$ and $V-I$ are plotted according to the $V$ stellar magnitude.
The observed change of color indices suggest for existence of a color reversal in the minimum light, a typical feature of the PMS stars from UXor type.
The result is evidence that the variability of GM Cep is dominated by variable extinction from the circumstellar environment.

\begin{figure*}
\includegraphics[width=5.5cm]{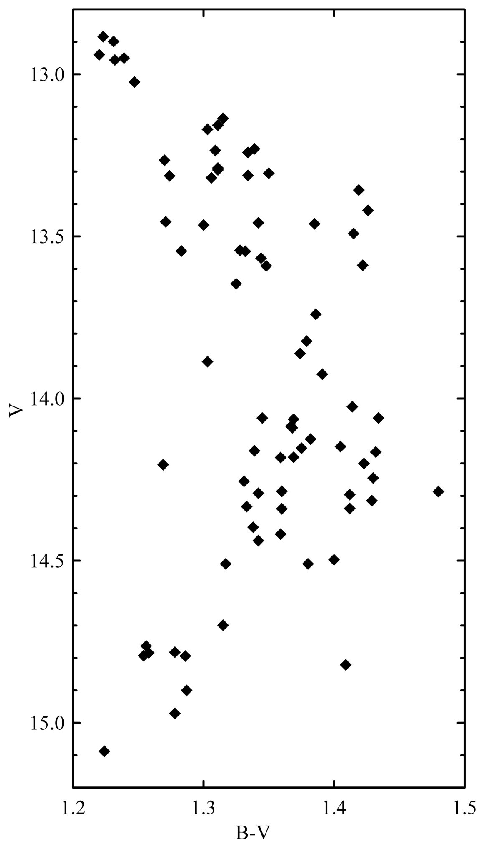}
\includegraphics[width=5.5cm]{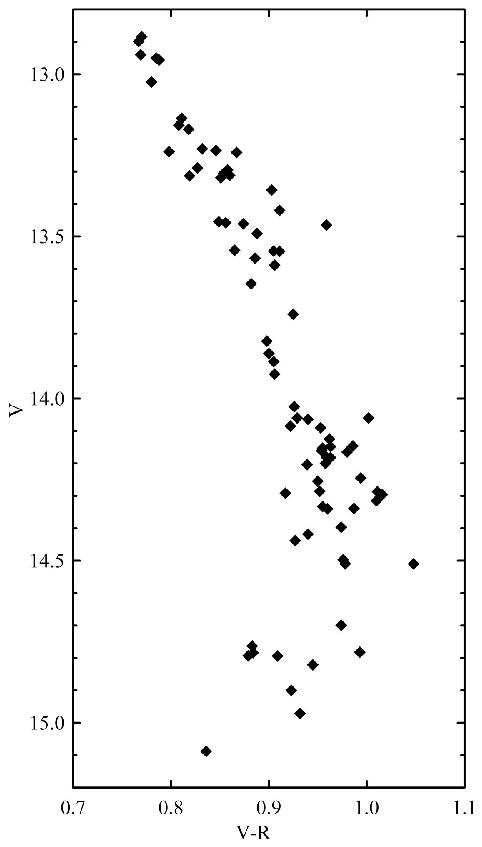}
\includegraphics[width=5.5cm]{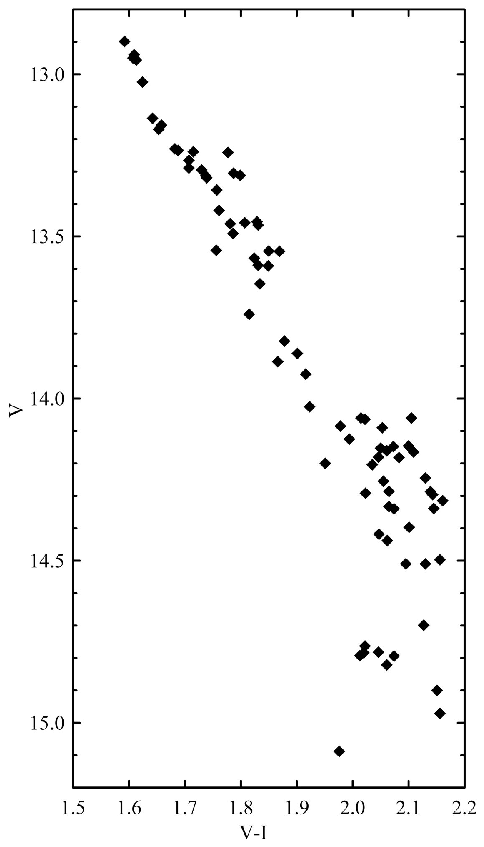}
\caption{The color-magnitude diagrames of GM Cep in the period of observations Jun. 2008 - Feb. 2011}
\label{figlabel}
\end{figure*}

The effect of color reversal (or co-called "blueing") has been studied and explained by many authors (Bibo \& Th\'{e} 1990, Grinin et al. 1994, Grady et al. 1995, Herbst \& Shevchenko 1999).
The widely accepted explanation of color reversal effect is variations of the column density of dust in the line of sight to the star, but the nature and origin of these dust clumps or filaments is still debated.
According to the model of dust clumps obscuration (an alternative model see in Herbst \& Shevchenko 1999) the observed color reversal is caused by the scattered light from the small dust grains.   
Normally the star becomes redder when its light is covered by dust clumps or filaments on the line of sight.
But when the obscuration rises sufficiently the part of the scattered light in the total observed light become significant and the star color gets bluer. 
Simultaneously with the decrease in brightness the percentage of polarization in the optical light increases dramatically (Grinin et al. 1994).
One of the first explanations of this phenomenon is that the stars are surrounded by number of protocometary clouds or cometary bodies (Grady et al. 2000 and references therein).
The more recent alternative model of Dullemond et al. (2003) proposes that UXor stars have self-shadowed discs, and the hydrodynamic fluctuations in the puffed-up inner rim of the disk can cause short time scale extinction events.
 
As we noted above GM Cep can be classified as ETTS, consequently it may show variability typical for both CTTS and HAEBE stars.
The broad H$\alpha$ emission line and the presence of a massive circumstellar disk are among the most important attributes of CTT stars. 
The strong P Cyg profile of H$\alpha$ line is interpreted by Sicilia-Aguilar et al. (2008) as evidence of very strong accretion.
Thus, the observed switching from high-brightness level to low-brightness level in the light curve of GM Cep (Xiao et al. (2010) can be explained by variable accretion rate typical of CTT stars (van Boekel et al. 2010).
This type of variability may explain also the rapid changes in brightness and the observed dispersion of the points on color-magnitude diagrams (especially on $V/B-V$).

The analysis of all collected data suggest that photometric properties of GM Cep can be explained by superposition of both: (1) magnetically channeled variable accretion from the circumstellar disk, and (2) occultation from circumstellar clouds of dust or from features of a circumstellar disk.  
Such a conclusion is mentioned in the paper of Sicilia-Aguilar et al. (2008), but the variable accretion is supposed to be the stronger contributor.
Our photometric results for the period Jun. 2008 - Feb. 2011 suggest that the variable extinction dominates the variability of GM Cep.
Analyzing the magnitude histograms from long-time light curves of GM Cep Xiao et al. (2010) reach a similar conclusion, the star spend most of its time in a nearly constant bright state, thus the variability is not dominated by flares caused by accretion.
In low accretion rates both types of variability can act independently during different time periods and the result is the complicated light curve of GM Cep.
Due to the complex circumstellar environment around PMS stars, such a mixture of different types of photometric variability can be expected.
A similar superposition of two types of variability is seen on the long-term light curve of another PMS star V1184 Tau (Semkov 2006, Semkov et al. 2008).
But in this case the observed light curve is due to the phenomena of both: the typical for WTT stars periodic variability (Tackett et al. 2003) and variable obscuration from UXor type (Barsunova et al. 2006).

\section{Conclusion}
At this stage it is most unlikely that the photometric variability of GM Cep can be attributed to outbursts from FUor or EXor type.
The long-time light curves of GM Cep show that the star spends much more time in high-brightness level than in low-brightness level.
Our study indicates that the high amplitude variability and the observed color reversal in the minimum light are caused by variable extinction from the circumstellar environment.
Simultaneously, brightness variability as a result of a variable accretion from the circumstellar disk are observed.

We consider the photometric studies of PMS stars as very important for their exact classification.
The problems from superposition of different types of variability can be solved by collecting of long-term photometric data from the photographic plate archives and from photometric monitoring in the present time.
Another disputed point that can be solved by both photometric and spectral monitoring is searching for correlation between the brightness of the star and the equivalent width of H$\alpha$ line.
Such a correlation between the deep minimums in brightness and the increasing of the equivalent width of H$\alpha$ line would be confirmation for the UXor type of variability of GM Cep.

\acknowledgements
      This work was partly supported by grants DO 02-85 and DO 02-273 of the National Science Fund of
      the Ministry of Education, Youth and Science, Bulgaria.
      The authors thank the Director of Skinakas Observatory Prof. I. Papamastorakis
      and Prof. I. Papadakis for the telescope time.
      This research has made use of the NASA Astrophysics Data System.

\end{document}